# Tackling Contention Through Cooperation: A Distributed Federation in LoRaWAN Space

Stéphane Delbruel
imec-DistriNet, KU Leuven
3001 Leuven, Belgium
stephane.delbruel@kuleuven.be

Nicolas Small
Optika Solutions Pty Ltd
nicolas.small@optika.com.au

Emekcan Aras
imec-DistriNet, KU Leuven
3001 Leuven, Belgium
emekcan.aras@kuleuven.be

Jonathan Oostvogels
imec-DistriNet, KU Leuven
3001 Leuven, Belgium
jonathan.oostvogels@kuleuven.be

Danny Hughes
imec-DistriNet, KU Leuven
3001 Leuven, Belgium
danny.hughes@kuleuven.be

## Abstract

Low-Power Wide Area Networks (LPWAN) play a key role in the IoT marketplace wherein LoRaWAN is considered a leading solution. Despite the traction of LoRaWAN, research shows that the current contention management mechanisms of LoRaWAN do not scale. This paper tackles contention on LoRaWAN by introducing *FLIP*, a fully distributed and open architecture for LoRaWAN that fundamentally rethinks how LoRa gateways should be managed and coordinated. *FLIP* transforms LoRa gateways into a federated network that provides inherent support for roaming while tackling contention using consensus-driven load balancing. *FLIP* offers identical security guarantees to LoRaWAN, is compatible with existing gateway hardware and requires no updates to end-device hardware or firmware. These features ensure the practicality of *FLIP* and provide a path to its adoption. We evaluate the performance of *FLIP* in a large-scale real-world deployment and demonstrate that *FLIP* delivers scalable roaming and improved contention management in comparison to LoRaWAN. *FLIP* achieves these benefits within the resource constraints of conventional LoRa gateways and requires no server hardware.

## 1 INTRODUCTION

Low Power Wide Area Networks (LPWAN) enable long-range sensing and control applications for the Internet of Things (IoT). LPWAN technologies such as LoRa[1], Sigfox[2], and NB-IoT[3] have achieved significant market traction and are rapidly being rolled out worldwide.

This paper focuses on LoRa, and its associated media access control protocol LoRaWAN, which together offer a set of attractive features, including: i) robust radio modulation, ii) an open protocol stack, and iii) no restrictions on the ownership and deployment of gateways. The LoRa protocol uses Chirp-Spread Spectrum (CSS) modulation to support the intermittent transmission of small packets according to a maximum duty-cycle that is defined by regional regulators such as the Federal Communications Commission (FCC)[4] and the European Telecommunications Standards Institute (ETSI)[5].

LPWAN networks, and LoRa in particular, are designed to support applications that demand low data rates, are delay tolerant and delivered using battery powered IoT devices. While LoRa may appear well-suited to this subset of IoT applications, there is growing concern over its ability to scale as the density of LoRa end-devices increases. The core of this problem is maintaining optimal throughput while an increasing number of end-devices perform unscheduled radio transmissions, that lead to collisions and packet loss. A growing body of research [2, 3, 23] shows that the scalability of LoRaWAN does not live up to the marketing claim that a single gateway can handle many thousands of end-devices [6].

This paper addresses the problem of contention on LoRa by contributing *FLIP*, the first fully distributed and open architecture for LoRaWAN gateways. *FLIP* securely federates LoRa gateways, which engage in coordinated load balancing to improve network reliability in dense deployments. This federated gateway architecture inherently supports the roaming of LoRa devices across the federation. *FLIP* achieves these goals, while preserving the same security and privacy features of the original LoRaWAN architecture. Evaluation shows that *FLIP* decreases channel utilisation by 45% in comparison to a state-of-the-art LoRa network server, while allowing 20% more devices to join. Furthermore, *FLIP* is fully backwards compatible with existing gateway hardware and requires no modifications to end-device hardware or firmware. This provides a feasible path to its adoption in the LoRaWAN marketplace.

[1] http://www.semtech.com/wireless-rf/internet-of-things/
[2] http://www.sigfox.com/en/
[3] http://www.3gpp.org/news-events/3gpp-news/1733-niot
[4] www.fcc.gov
[5] www.etsi.org

The remainder of this paper is structured as follows: Section 2 provides an in-depth analysis of the factors that limit LoRaWAN scalability along with requirements for addressing these problems. Section 3 describes the *FLIP* architecture. Section 4 presents its implementation. Section 5 then evaluates *FLIP* in a large-scale real world deployment. Section 6 reviews prior work on improving LoRaWAN scalability. Finally, Section 7 discusses directions for future work and Section 8 concludes this work.

# 2 BACKGROUND

The introduction of LPWAN technologies such as LoRaWAN, SigFox and NB-IoT has generated significant interest in the IoT community, due to the promise of economically delivering long range, low power networking.

The LoRaWAN network protocol stack is based upon the LoRa physical radio layer originally developed by Cycleo and later acquired by Semtech. LoRa is an open protocol, the development of which is guided by a consortium composed of private and public actors[6]. In contrast to SigFox and NB-IoT, LoRa users are free to independently deploy LoRa networks and this free approach to deployment has gained LoRaWAN significant market traction. However, the rapid global roll-out of uncoordinated LoRa networks is testing the fundamental scalability of the protocol and a growing body of research now suggests that LoRaWAN will not scale to support dense networks [2, 3, 23].

This remainder of section analyses the LoRaWAN protocol and identifies key scalability concerns. Based upon this analysis, we then review existing approaches to improve LoRaWAN scalability and make the case that LoRa's scalability problems can be addressed most effectively through cooperation and federation.

## 2.1 Scalability limitations in LoRaWAN

The scalability limitations of LoRa and LoRaWAN arise from multiple sources at the Physical (PHY) and Medium Access Control (MAC) layers as well as the distributed system architecture of the LoRa backhaul network that links gateways to the network servers that control them. For brevity, we do not provide a complete description of the LoRaWAN stack. For a detailed review of the protocol we refer the reader to the LoRaWAN technical specification [17].

### *2.1.1 Physical layer limitations*

The LoRa physical layer operates in the unlicensed sub-GHz frequency band with specific frequency ranges defined according to regional regulations. LoRa is a derivative of Chirp Spread Spectrum (CSS) modulation technique, to transmit information. The spreading of the spectrum is achieved by generating a chirp, a signal in which the frequency linearly increases *up-chirps* or decreases *down-chirps* over time, achieved by modulating the phase of an oscillator over a certain bandwidth. Chirps are generated by changing the carrier phase of the transmitter in accordance with a binary sequence, where each basic element is called a chip, to avoid confusions with the *bits* used in the data sequence. The data sequence is multiplied by this chip sequence and modulated onto the chirp signal making CSS a robust technique for the long range transmission of data. LoRa enables developers to trade off between range and throughput by appropriately configuring a *Spreading Factor* parameter that determines the number of chips in a symbol. LoRa defines six spreading factors that range from SF7 to SF12 where each incremental increase in SF doubles the symbol duration and therefore the time that is required to transmit a packet of fixed size. The transmission time for a 1 byte LoRa packet ranges from under 100ms at SF7, with a theoretical range of under 1km to over 1.8s at SF12, with a theoretical range of over 20km.

High SF transmissions that occupy the shared network medium for lengthy periods across a long range lead to a clear *contention* problem. The LoRa physical layer provides three countermeasures against this problem. First, the use of CSS modulation enables LoRa gateways to receive multiple messages simultaneously if those messages are transmitted at a different SF. CSS cannot prevent partial or full packets losses due to contention in cases where the colliding messages share the same SF even if received with a sufficient power differential. Second, LoRa gateways offer up to eight upstream channels, which enables packets to be received simultaneously on orthogonal sub-bands within the LoRa frequency space. Third, regional LoRa duty cycle limitations constrain how frequently end devices may send packets, reducing contention by limiting throughput.

While the techniques described above incrementally address the scalability limitations of LoRa, they are inherently limited by the underlying LoRa MAC protocol, which implements an ALOHA-like uncoordinated medium access scheme, a technique which is known to lead to very poor efficiency in dense deployments. This theoretical observation is validated by empirical research, which shows that LoRa does not scale to support dense deployments [9, 20, 5, 10]. We discuss this problem in the following sub-section.

### *2.1.2 Medium Access Control layer limitations*

LoRaWAN protocol specifies three classes of devices:

- *Class-A devices* transmit messages upstream to the gateway in an uncoordinated fashion and only listen for downstream messages for a brief period after the upstream transmission. Downstream messages are queued at the gateway until such a downstream transmission slot occurs.
- *Class-B devices* synchronize their internal clocks with a beacon that is provided by the gateway and, based upon that signal, negotiate scheduled downstream transmission slots. This approach enables lower latency downstream transmissions to support actuation applications. Upstream messages are handled in the same was as Class-A devices.

[6] https://www.lora-alliance.org

- *Class-C devices* send upstream transmissions in the same way as Class-A devices, however, they listen continuously for downstream messages, minimising downstream latency at the expense of power consumption. This mode of operation is expected to be deployed in powered scenarios.

The upstream behaviour of all classes of LoRa devices closely approximates that of the well-known ALOHA protocol, including the inability to detect collisions during transmission due to the lack of separate upstream and downstream channels. It is well known that ALOHA-like MAC protocols result in poor channel utilization due to collisions resulting from uncoordinated transmissions [1].

LoRaWAN provides a number of remote MAC commands through which gateways may configure end-devices. Critically for managing contention and interference, LoRa supports Adaptive Data Rate (ADR), which enables a gateway to command an end-device to switch data rates, transmit power, and channels, so as to optimize use of the shared network channel. However, research has shown that ADR performs poorly in cases of contention between collocated networks [24, 21].

#### 2.1.3 LoRa gateway and backhaul limitations

The backhaul architecture of LoRaWAN is composed of a number of gateways, which relay messages from the LoRa network to a single coordinating *Network Server*. This architecture is referred to as a *star of stars*. Each LoRa gateway may provide an upstream link for many end devices and it is possible that messages from one end-device may be received and forwarded by multiple gateways. Downstream messages on the other hand are transmitted to end-devices via only one gateway that is selected by the network server.

The backhaul architecture of LoRa provides no mechanisms for coordinated optimization across co-located LoRa networks that are operated by different actors. This lack of cooperation between network operators may cause end-devices to be optimized in a way that negatively impacts co-located networks. The result is that the performance of a LoRa network will vary unpredictably based upon the networks that are deployed around it.

In cellular long range networks, these problems have been addressed through a rich set of inter-cell coordination and management protocols, together with extensive cooperation and federation between network operators. However this form of heavyweight coordination and planning is antithetical to the free deployment philosophy of LoRaWAN. The following subsection discusses how the LoRaWAN community have responded to tackle the problem of contention.

### 2.2 Known Approaches to Scaling LoRaWAN

The LoRaWAN community is aware of the scalability problems facing the protocol. Attempts to address the problem fall into three broad categories: increasing gateway density, developing new MAC protocols and improved use of existing ADR commands. We briefly review each stream of work in the following subsections.

#### 2.2.1 Building dense gateway deployments

The most obvious way to improve LoRaWAN performance is to increase capacity by deploying additional gateways. By reducing the average distance between gateways and end-devices, Spreading Factors can be reduced through standard ADR mechanisms so that average transmission times and therefore contention are minimised. In cases where directional antennas are used intelligently, significant gains also may be possible [26] at the expense of greater deployment complexity.

Deploying dense gateway infrastructure has a number of shortcomings. Firstly, this approach increases costs and complexity for the end-users who must deploy additional hardware, power and backhaul network connections. Secondly, dense gateway deployments erode the core value proposition of LoRa: economic long range networking. It is not clear where this trend will end. For example, at SF7, the contention problem of LoRa is much reduced, but effective range falls to several hundred meters, at which point the protocol's competitiveness against short range network technologies such as IEEE 802.15.4, WiFi or BLE becomes unclear.

#### 2.2.2 New MAC protocols

The development of more scalable MAC protocols for LoRa holds great potential. It is well known that the ALOHA-like MAC protocol of LoRa cannot scale to support dense networks. Moving from the current unsynchronized protocol to a time-slotted protocol, or better yet a time synchronized MAC protocol should be expected to provide a significant performance improvement.

The research community has contributed a number of promising candidate protocols to address this challenge such as [12]. However, deploying these proposals will be difficult as they require a wholesale reworking of the LoRaWAN network architecture. Furthermore, new MAC protocols do nothing to improve performance for the millions of LoRaWAN devices that are already deployed today, and that the new generation of MAC protocols will be forced to contend with for years to come.

#### 2.2.3 Better use of ADR commands

An orthogonal stream of research aims to make the best possible use of the current LoRaWAN MAC protocol by optimizing or redesigning the Adaptive Data Rate (ADR) protocol. Research in this area has identified a number of shortcomings in the design of ADR [24, 21, 19] and complementary techniques for improving the protocol have been proposed [25].

The advantage of these proposals is two fold. First, they are easy to deploy as they require no modifications to end-devices firmware or low-level protocol elements. Second, better ADR schemes can improve performance for the millions of LoRa devices that are already deployed in the field. Unfortunately, it seems unlikely that ADR optimization alone can deliver sufficient performance improvements as ADR cannot address the problem of contention between co-located networks. While version 1.1 of the LoRa specifications [18] introduced limited support for gateway coordination in the form of *passive roaming* and *handover roaming*, there is, to date, no support for distributed optimization of gateway access.

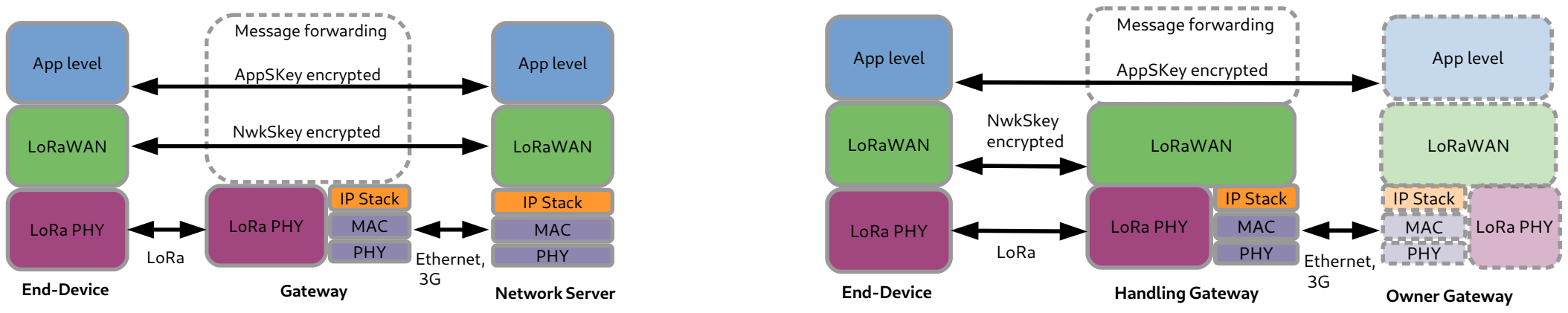

**(a) Current centralized interactions model**

**(b) *FLIP* distributed interactions model**

**Figure 1: *FLIP* places the NetworkServer role inside the gateways and allows the role of the handler to be decoupled from the role of the owner if necessary.**

## 2.3 Requirements

LoRaWAN faces significant performance challenges that arise due to contention as described above. In our view, a solution to this problem must respect the following three requirements:

1. *Maintain backwards compatibility with end-devices*: Millions of LoRa devices are already deployed in the field and may remain operational for many years to come. Effective approaches to improving LoRaWAN performance should co-exist with deployed LoRaWAN v1.0 devices or, better still, effectively improve the performance of these devices.

2. *Encourage cooperation instead of competition* between co-located LoRa gateways in order to make the most of existing back-haul capacity and reduce costs that would otherwise arise due to the deployment of redundant gateway equipment and additional coordination servers.

3. *Preserve flexible gateway deployment*: A major strength of the current LoRa network, is that end-users are free to deploy gateways and set up private networks without permission or coordination. Any approach to addressing the scalability limits of LoRaWAN should respect this deployment philosophy by providing end users with equivalent device management features and ensuring that unreasonable coordination burdens are not passed on to network operators.

4. *Robust scalability*: A solution to LoRa's scalability issues should deliver a robust system architecture that avoids central points of coordination and therefore failure. Furthermore, the architecture must robustly handle the dynamics of LoRa networks such as churn, load dynamics and mobility.

This paper argues that these requirements can most effectively be met by transforming LoRa gateways from centrally controlled packet forwarders into first class members of a federated network ensuring scalability by maximizing cooperation between LoRa networks, on the rationale presented in Figure 1. In the following section, we introduce the architecture of *FLIP*, designed to address these requirements.

# 3 SYSTEM ARCHITECTURE

The design of *FLIP* is based upon the core observation that for each additional LoRa device added to the network, there is a geographically limited increase in contention. This increase in contention can most appropriately be tackled by coordinated action taken by co-located gateways. The remainder of this section is structured as follows. Section 3.1 provide a high-level overview of the FLIP architecture. Section 3.2 describes local clusters, the basic unit of organization in FLIP. Finally, Section 3.3 describes how local clusters coordinate to form a large-scale federation.

| DevEUI | RSSI |
|---|---|
| 2F0C5E21DFB124E6 | -80dBm |
| A1C32AD2F275E6B1 | -102dBm |
| 68A378D591FC1128 | -57dBm |
| ... | ... |

**Figure 2: Gateway profile: Each gateway completes its own profile with observed end-devices under the form of a two dimension vector. For each observed device, DevEUI and associated signal strength compose direction and magnitude.**

## 3.1 Architectural Overview

The design of *FLIP* embodies three main architectural principles. First we ensure scalability through a hierarchical approach that mirrors the principles of the Internet architecture. Second, we reduce the perimeter of coordination to the greatest extent possible using locality-aware communication. Third, we enable cooperation with consensus-based decision making. These principles are explored in Section 3.1.1 to 3.1.3 respectively.

### *3.1.1 Scalability through hierarchy*

*FLIP* connects all participating gateways at the IP level using a decentralized Virtual Private Network (VPN). This federation-wide VPN forms the basis for all higher level communication. In order to maintain scalability, *FLIP* follows the same hierarchical model that is used in the Internet. Incoming gateways join a *local cluster*, which supports localised data exchange as described in Section 3.1.2. Each local cluster is assigned a unique IP range, within which standard network-level broadcast can be used to discover fellow cluster members. Each cluster autonomously elects a temporary *cluster leader*, who has the additional responsibility of building routes to the remote clusters in the federation. This is analogous to the way in which border routers connect autonomous systems in the Internet architecture. The cluster elects a leader based upon the consensus protocol described in 3.1.3.

### *3.1.2 Localised and information flows*

Local clusters are the basic unit of organization in *FLIP* and may range in size from a few gateways deployed in users homes to several thousand gateways that coordinate to provide seamless network coverage for a smart city. Information flows within the cluster occur most frequently between gateways that are close to each other in radio space. *FLIP* exploits this communication pattern by building a proximity-based overlay network for data dissemination. Cluster members periodically exchange *node profiles* that contain a list of visible end devices, along with RSSI data. Adding along *Signal-to-noise ratio* (SNR) to the profiles have been considered and dismissed due to limited added value compared occurred additional network costs. These profiles are considered as multidimensional vectors, where the unique identifier of one received end-device represents one dimension and it's associated signal strength its norm, as depicted in Figure 2. Proximity is calculated using a distributed $k$ Nearest Neighbour ($k$NN) algorithm supported by a cosine similarity classifier [8]. Based upon this proximity information, *FLIP* builds an application-level overlay [14] that mirrors the distribution of cluster members in radio space. Local maximum issues are avoided using a Random Peer Sampling (RPS) service [13]. Information is disseminated across this overlay using a gossip protocol [11, 14], which ensures a bounded worst-case load and hence guaranteeing scalability.

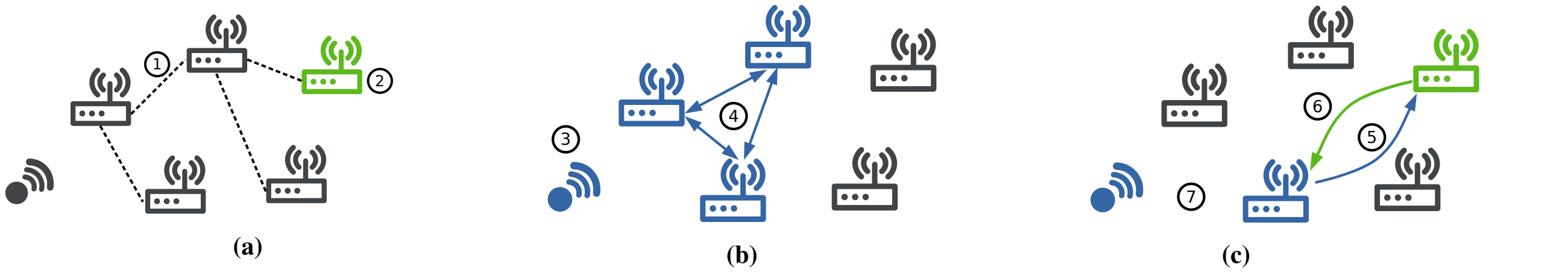

**Figure 3: Interactions inside a *FLIP* cluster are composed of three stages: a) Welcoming a new actor and integrating it to the cluster self-organization process. b) Welcoming an end-device and determine collaboratively how to absorb this new load the most balanced way possible between participating members. c) Having the best suited actor negotiates the handling delegation with the owner of this device in order to operate it and forward the produced data.**

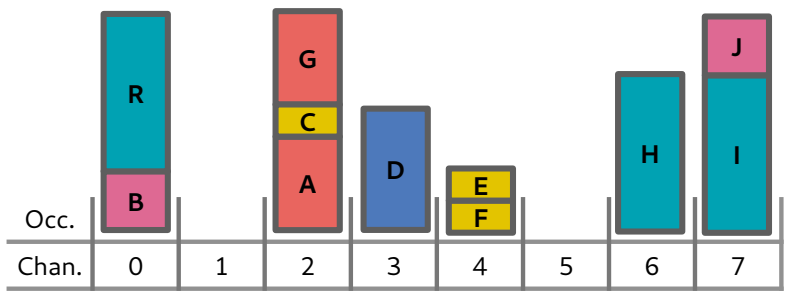

**Figure 4: Channel occupation: Attaching end-devices (represented here by letters) to a gateway increases the joined channel occupation by their respective radio resource consumption (here block size). For similar payload lengths and duty-cycle, the additional load in radio resource consumption is a function of their respective *SpreadingFactors* (each corresponding here to a colour).**

#### *3.1.3 Cooperation through consensus*

*FLIP* is self-organising and when decisions must be taken within a cluster, they occur through consensus. This applies to two key operations: the election of cluster leaders for inter-cluster routing as described in 3.1.1 and the election of a handler gateway that will accept the traffic of an incoming LoRaWAN end device. In both cases, we use the asynchronous byzantine leader election method of Kapron et al. [15] which is fast and fault tolerant. Handlers ensure good use of spectrum by measuring the occupation of their channels and those of its neighbours. Channel occupation is defined by the sum of the weight of each end-device present on a given channel, where the weight is determined by its radio resource consumption, as a function of the set transmit duty-cycle, *SpreadingFactor* and payload length. Multiple device occupying the same channel of their joined gateway add their respective weight to the overall channel occupation as in Figure 4. Incoming devices are then allocated to the least occupied channels among the co-located nodes in order to evenly distribute channel occupation across all gateways in range, according to the Shannon entropy measure:

$$H(X) = -\sum_{i=1}^{n} \mathrm{P}(x_i)\log_b \mathrm{P}(x_i)$$

The following sub-sections detail how the different elements of the *FLIP* architecture interact to provide a seamless backhaul for the LoRa network. Section 3.2 describes local cluster interactions, while Section 3.3 describes federation-wide interactions

### 3.2 Local cluster interactions

Participation in the local clusters is comprised of three phases of operation outlined below and pictured in Figure 3.

***Connecting a gateway to the federation:*** An actor that wishes to participate in the *FLIP* federation deploys the *FLIP* software on their gateway and uses this software to connect to the local cluster of their choice via any of its current cluster members. Once a gateway is connected to the cluster, it exchanges its *profile* with the other cluster members via gossip, becoming a part of the proximity-aware overlay network as described in Section 3.1.2 and depicted as ① in Figure 3a. Once connected, the proximity graph is periodically refreshed as new profiles are exchanged.

***Initiating end-device connection:*** A participant that wishes to deploy an end-device must first advertise it by propagating an ownership marker from their gateway across the proximity graph, ②. The ownership marker is comprised of the device unique identifier (devEUI), the identity of the owner and period of ownership validity. Once ownership is established, an end-device may attempt to join the *FLIP* federation, by transmitting a JoinRequest using the standard LoRaWAN Over-The-Air-Activation protocol, ③ in Figure 3b. Any federation member that receives this join request will first check its ownership database for a matching marker. If found, candidate gateways will initiate the distributed consensus algorithm described in Section 3.1.3 to elect a handler for the incoming node based upon a measure of their entropy ④.

***Delegating device management:*** Once the join process has been initiated and consensus reached, the elected handler will then contact the owner of the incoming end-device to initiate the delegation process, ⑤ in Figure 3c. The end device owner then generates the standard set of device keys, sending the join acknowledgement back to the end device via the handling gateway (6). The device owner then shares the NetworkSessionKey with the handler to delegate management of the end-device while keeping the ApplicationSessionKey private. This ensures end-to-end privacy between the incoming end-device and the owning gateway. The join procedure is completed when the handler sends a *JoinAck* to the device along with the selected network settings (7).

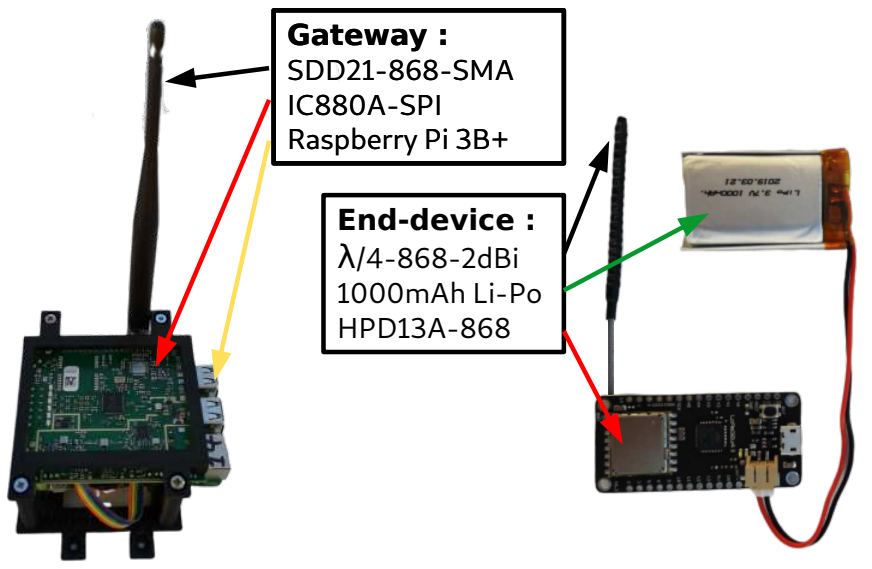

**Figure 5: Hardware setup: Depicts LoRaWAN gateway and client end-device embedding a temperature sensor.**

## 3.3 Inter-cluster interactions

The participation of the cluster leaders in routing across the federation has three key elements, as described below.

***Electing a cluster leader:*** The first step is for each cluster to periodically elect a subset of its members to the position of *cluster leader* using the local consensus algorithm that is described in Section 3.1.3. The current election process takes into account the relative resource utilisation of each gateway, electing the leader that will maximally balance load.

***Discovering remote clusters:*** The cluster leaders execute a peer discovery protocol to discover distant cluster leaders and, in turn, notify those gateways of their presence. Based upon the remote cluster leaders that are discovered, each cluster leader builds an inter-cluster spanning tree that connects the remote clusters via their leaders using the well known link state routing approach[7].

***Routing across clusters:*** Cluster leaders share ownership information between each other using the same mechanism as is applied within a cluster. This causes end device ownership to propagates across the federation. When an inter-cluster ownership message reaches a *cluster leader* it will be propagated along the inter-cluster spanning tree. Each leader that receives this message will then initiate a local dissemination of this marker in its respective cluster, before forwarding it on to the rest of the federation.

# 4 IMPLEMENTATION

To ensure the reproducibility of our results and to foster research collaboration we have published the complete implementation of *FLIP* as free and open source software [7]. At the physical level our implementation builds upon Semtech's packet forwarder[8], while at the network layer *FLIP* builds upon Tinc[9], a self-routing mesh-networking protocol used for private network service, which provides gateways with virtual IP addresses. For the purposes of our experiments, we chose to allocate each possible cluster a /24 IPv4 subnet, allowing for up to 254 gateways per cluster. Any underlying network can provide inter-gateway connectivity. In the case of our experiments, we use Internet via standard ADSL.

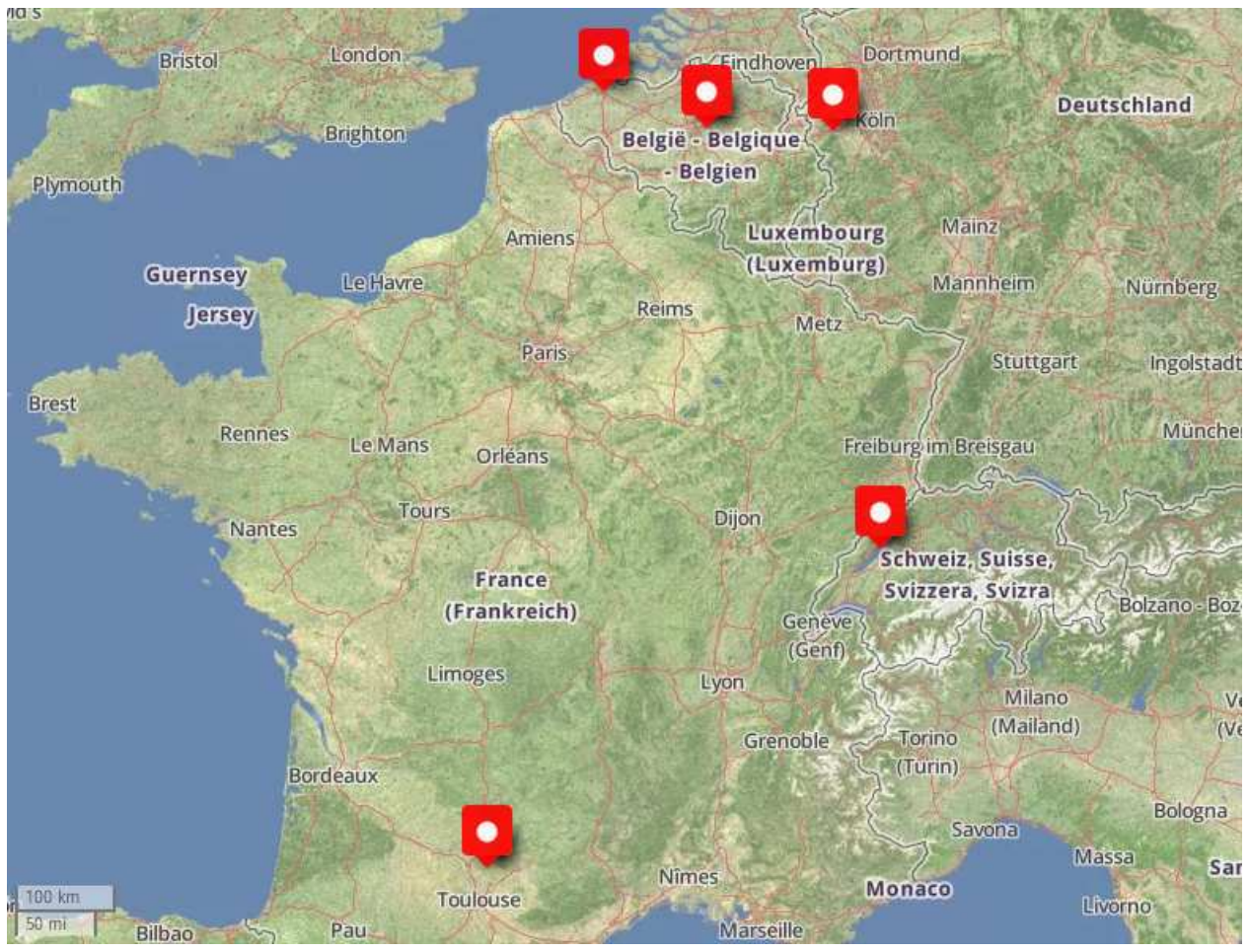

**Figure 6: Presence of *FLIP* clusters in Europe across 4 countries via 5 cities.**

In order to encourage end-user adoption and participation in the *FLIP* federation, we have also released a Raspbian[10] image that bundles all necessary elements into a single package [7]. For the purposes of this paper, the image is used to operate Raspberry Pi-based gateways that are connected to an iC880A LoRa concentrator via Serial Peripheral Interface (SPI). These gateways are assembled as in Figure 5, and connected to the Internet over standard Ethernet. End-users can replicate this setup using the provided image and the associated commercial off-the-shelf hardware. This enables the user to register end-devices which can freely roam across the federation by simply plugging in an Ethernet cable and configuring their own gateway for a ~200€ price point. The end-devices used for the deployment and the evaluation are also presented in Figure 5 and composed of an ATMEGA32u4 with an internal temperature sensor, powered by a Li-Po battery of 1000mAh, connected to a HPD13 LoRa module transmitting at a constant power of +14dBm via a 2dBi quarter-wave monopole antenna with a SWR of 1.108 at 868MHz. These end-devices can be sourced[11] at a 13€ price point.

# 5 EVALUATION AND PERFORMANCE

Over the course of six months, we have established a testbed to support field trials of *FLIP* by deploying 17 *FLIP*-enabled gateways along with 417 end-devices across 5 cities located in 4 different countries: Brugges (BE), Leuven (BE), Aachen (DE), Neuchâtel (CH) and Aveyron (FR) as illustrated in Figure 6. Each city hosts a single local cluster which contains a variable number of gateways and end-devices.

This multi-country deployment is used to support a two-phase evaluation, which assesses key dimensions of perfor-mance. Section 5.1 provides the technical details of the experiment setup. Sections 5.2 to 5.4 evaluate the ability of *FLIP* to tackle contention within a single cluster, while Section 5.5 then examines the scalability of *FLIP* at the level of individual gateways and the federation as a whole.

[7]https://tools.ietf.org/html/rfc3626
[8]https://github.com/Lora-net/packet_forwarder
[9]https://tinc-vpn.org/
[10]https://www.raspberrypi.org/downloads/raspbian/
[11]https://bsfrance.fr

| SF | 7 | 8 | 9 | 10 | 11 | 12 |
|---|---|---|---|---|---|---|
| **ToA** (ms) | 46.35 | 92.67 | 164.86 | 329.73 | 659.46 | 1155.07 |

**Table 1: PHY payload Time-on-Air for various SFs.**

To the best of our knowledge, this work is the first to evaluate contention with multiple gateways and hundreds of end-devices in a real world deployment environment. As such, there is no directly comparable work that we can evaluate against, so we instead compare our results to the performance of a standard LoRa network running on identical hardware. To ensure a fair comparison, we evaluate in harsh conditions, where our baseline already reach the maximum theoretical predicted performance of LoRa.

## 5.1 Experimental Setup

Our evaluation is comprised of two phases, the first focusing on how our work helps tackle the contention issues caused by non-cooperative actors, the second demonstrating the scalability and thus validity of our solution. All experiments were conducted using the common hardware base described in Section 4. In the interests of transparency, all of the necessary software and configuration files required to reproduce the experiments have been published online [7].

We begin in Section 5.2 by focusing on the contention handling performance of *FLIP*in comparison to uncoordinated LoRa gateways, which are gradually connected to the federation wherein they are coordinated by *FLIP*. The firmware variants are executed on the same physical gateways in order to ensure consistent hardware and network characteristics across the various experiments. We deploy 4 gateways across a semi rural area in Aveyron, France, in range of each others and introduce 417 end-devices in their coverage area. Mirroring the abstract model in [26], the geographic positioning of these gateways denoted $A$ to $D$ is presented in Figure 7 and is centred on gateway $D$. The 417 LoRaWAN end-devices are distributed among the population of this village to be placed in houses, public buildings and warehouses. They transmit a fixed size payload containing a temperature measurement every 3 minutes, allowing the devices operating on SF12 to be compliant with duty-cycle limitations of 1% while maximizing the resulting contention. The radio settings of each end-device are fixed by the gateway during the join request with $CF = \{867.1; 868.5\}$MHz while the Spreading Factor is decided by the end-device as the minimum needed to join the network with $SF = \{7; 12\}$. Bandwidth settings are fixed at 125kHz along with a join cycle period of 8 hours. Upstream messages follow the standard format according to specifications, have a fixed application payload of 2 bytes, which results in a fixed PHY payload length of 15 bytes with the associated time on air of these packets at different spreading factors indicated in Table 1. Our chosen deployment conditions are harsh, increasing contention and thus demonstrating the capability of *FLIP* to increase network performances even in unfavourable conditions. The present settings will generate one *JoinRequest* every 70 seconds on average.

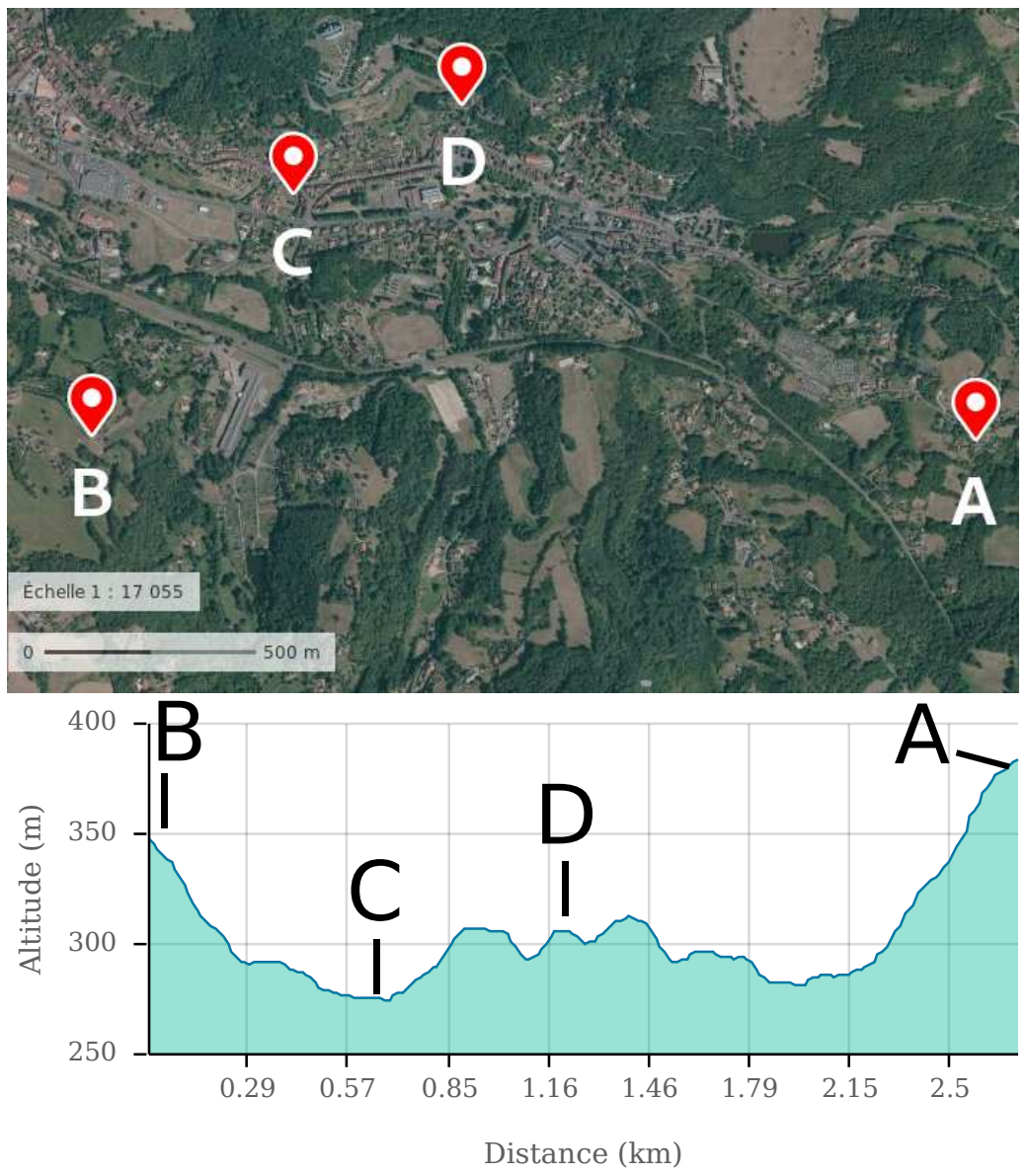

**Figure 7: Urban deployment of gateways used in contention evaluation and its corresponding altimetric profile.**

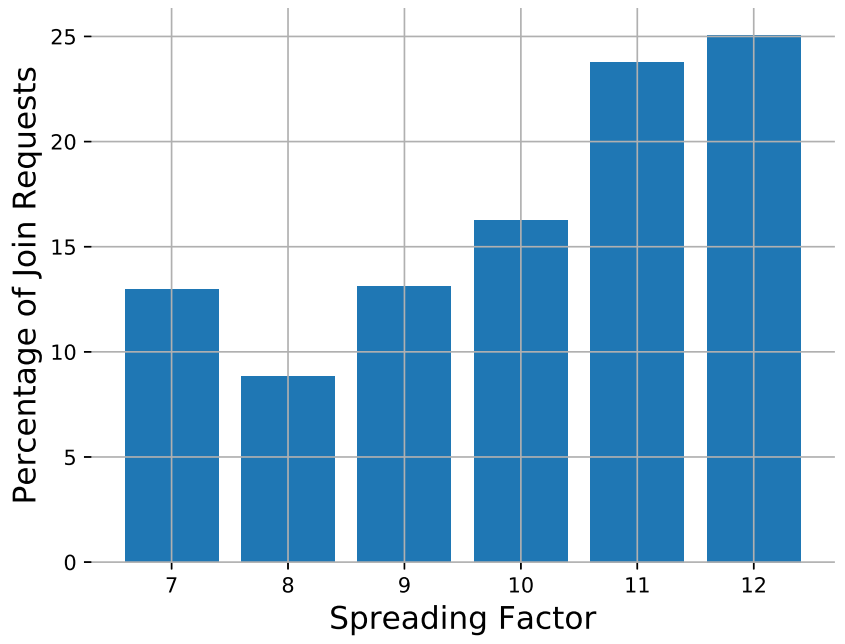

**Figure 8: Assessing the radio environment - Distribution of Spreading Factors for transmitted *JoinRequests* in a single owner case without interfering networks.**

As the corresponding *JoinAccept* emitted by the gateway is 23 bytes long, and the gateway is also subject to 1% duty-cycle limitations, a gateway will be able to handle all the devices if no transmissions collide and all devices join with a *SpreadingFactor* of 10 or lower, which is made unrealistic by the nature of our deployment, the locations of end-devices and their distance to the gateways, thus degrading the radio environment and maximising the generated contention.

## 5.2 Handling Devices from a Single Owner

The following scenario assesses the performance of *FLIP* in a scenario with low contention, where all devices present are provided by a single owner in the absence of competing networks. This allows us to study the unique conditions in which our subsequent experiments are performed. The more devices that a gateway handles, the more contention occurs during device transmissions. This impacts both normal traffic and the LoRa join process, potentially preventing additional nodes from joining.

In this first scenario, gateways $A$, $B$ and $D$ are active along

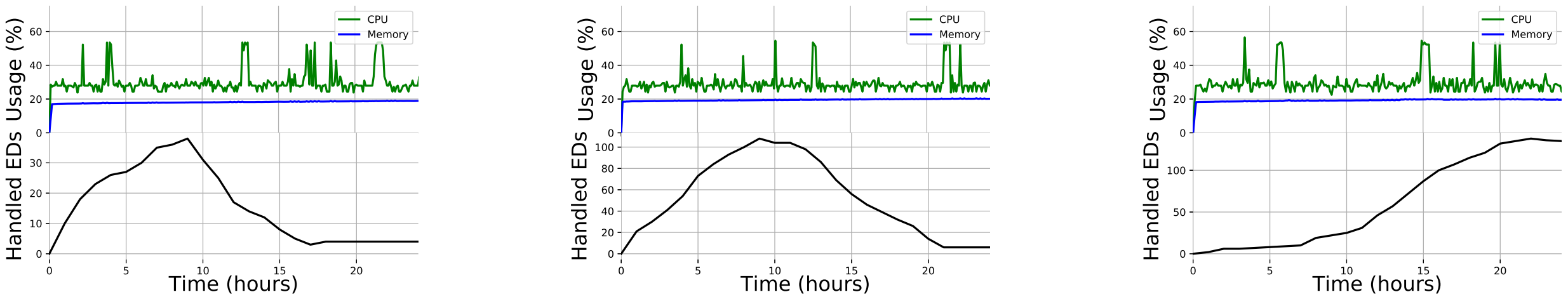

**Figure 9: Cost of *FLIP*: Gateways *A* (left) and *B* (centre) work with gateway *D* (right) to handle its devices in an efficient manner. Regardless of how many end-devices each gateway handles, CPU and Memory costs for each gateway remain constant.**

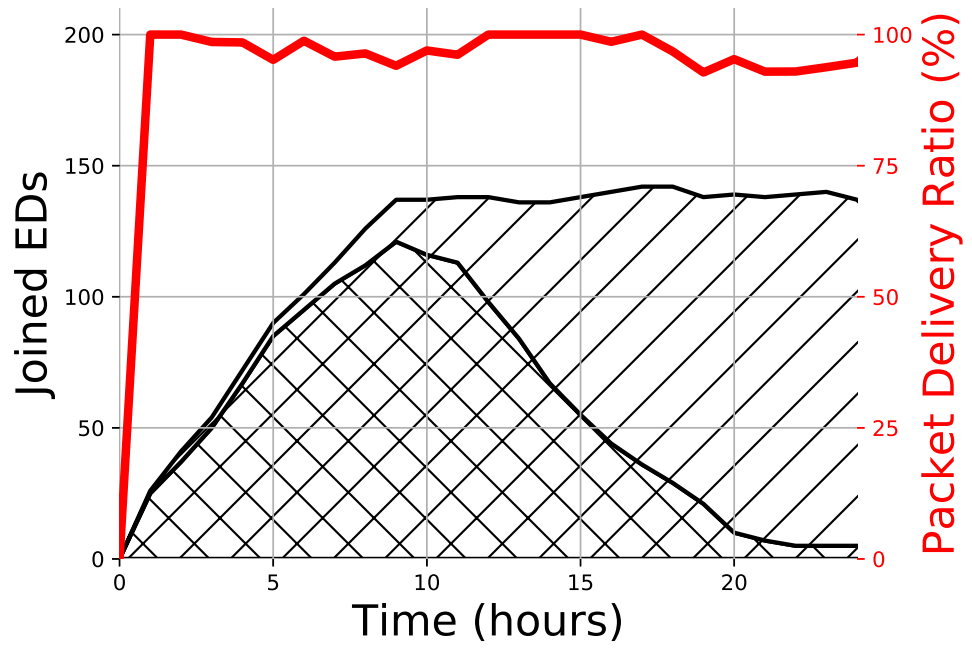

**Figure 10: PDR stays above 90% regardless of how many end devices join, whether these devices are handled directly by the owner (hatched) or delegated (cross-hatched)**

with 300 end-devices belonging to *D*. All devices are deployed in the vicinity of *D*, but within range of *A* and *B*. Each device attempts to join its desired gateway at the lowest spreading factor. If the join attempt is unsuccessful, the device increases its spreading factor incrementally and retries until it succeeds in establishing a connection to the gateway. In this way, spreading factors are minimized to the greatest extent possible.

The distribution of the transmitted JoinRequests and their respective SFs is an indicator of how end-devices have to compete to join the network. To join the network using Over-The-Air Activation (*OTAA*), a gateway must be able to receive the corresponding *JoinRequest* and respond with a *JoinAccept* in a timely manner while respecting the downstream duty-cycle restrictions that apply to the gateway. As such, the percentage of end-devices that are able to join the infrastructure and dully operated by their respective owners is directly influenced by radio conditions. In this case, the distribution of end-devices requires a significant proportion of high Spreading Factors in *JoinRequest*, as visible in Figure 8. This is due to contention at join time, which forces end-devices to perform multiple successful *JoinRequest* before being handled.

Figure 9 shows how the three gateways handle devices over a 24 hour period wherein each end-device performs at least 2 join cycles. For each gateway, we show how the number of handled end-devices evolves along the associated performance costs of *FLIP* for these cooperating gateways in terms of CPU and memory resources consumed by that collaboration in Figure 9.

***Delegation mitigates join-time contention:*** As nodes begin to join the network, actor *D* becomes overloaded, being unable to complete the OTAA process in a timely fashion due to LoRa duty cycle limits. Thus actors *A* and *B* handle the majority of the end-devices in the first 8 hours of the experiment. As these nodes are more distant from *A* and *B*, they are forced to join the network at higher Spreading Factors. As the join time contention on *D* abates, actors *A* and *B* cede control of these devices back to actor *D*, which enables the end-devices to reduce their Spreading Factor and thus operate more efficiently.

***The overhead of FLIP remains low:*** It is important to note that there is no correlation between the CPU usage and memory and the number of end-devices that a gateway must handle. A small increase in memory consumption occurs during the initialization phase of the experiment, but memory use then remains stable throughout the experiment. The spikes evident in CPU usage arise due to periodic .zip compression of syslog files and experiment logs, which relates to the operating system and the experiment respectively instead of to *FLIP* itself.

***FLIP ensures a high packet delivery ratio:*** The coordinated delegation of devices across gateways maximises the overall Packet Delivery Ratio (PDR) of the network as shown in Figure 10. Regardless of the increase in the handled end-devices or the delegation process, the overall PDR of that stays consistently between $95-100\%$, which is superior to the simulated performance predicted in [5].

Considered in sum, the delegated handling of end-devices reduces join time contention, ensures a high packet delivery ratio and incurs limited memory and computational overhead on the participating gateways.

## 5.3 Handling Devices from Multiple Owners

We previously illustrated how *FLIP* ensures collaboration between multiple gateways in the absence of interfering networks. This second scenario examines the performance of *FLIP* when the ownership of the previous 300 end-devices is shared equally among the participating actors *A*, *B* and *C*. This is compared against three uncoordinated networks. The end-devices are once again deployed in close vicinity of *D* but now belong to 3 different actors. While the devices are closer to *D*, they are within range of their owning gateway.Due to the greater distance between end devices and their owning gateway, we observe a greater proportion of high spreading factors in the *JoinRequests*, with a notable 34% transmitted at SF12 as shown in Figure 12. The use of higher spreading factors increases time on air and therefore contention [16].

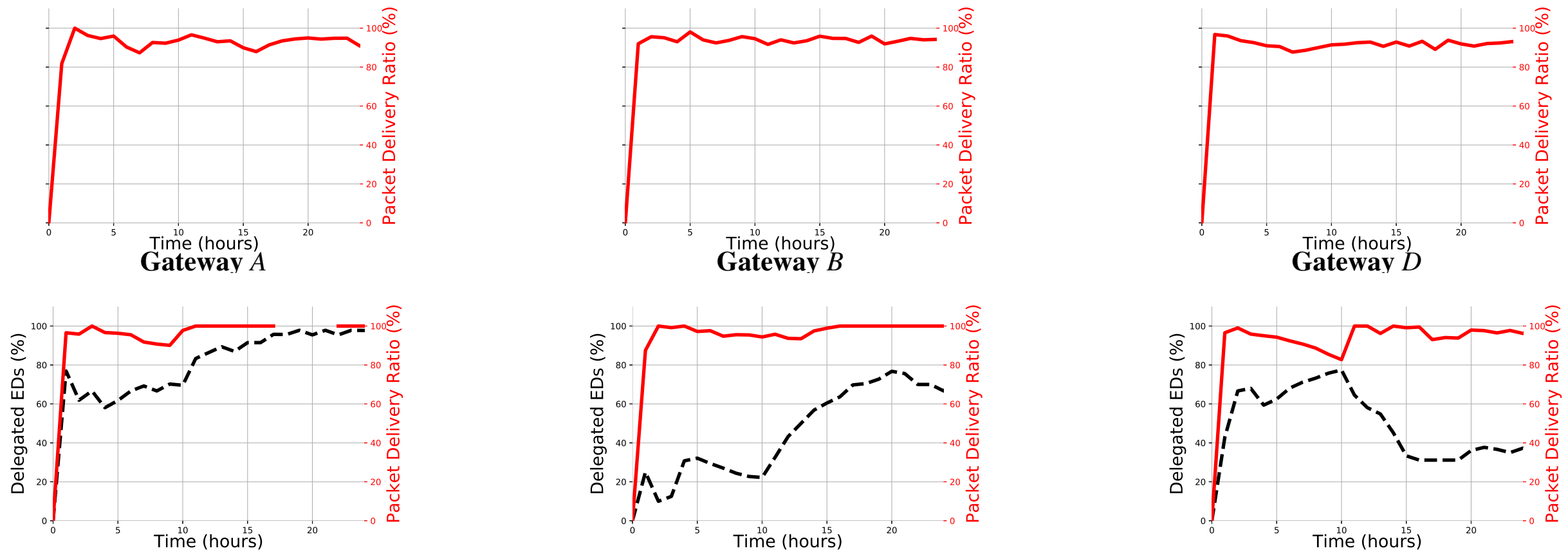

**Figure 11: Tackling distance generated contention: Respective PDRs of gateways $A$, $B$ and $D$ when acting as independent actors (top), and when collaborating through *FLIP* (bottom). Collaboration context is provided for each gateway via the ratio of owned end-devices handled by another collabor**

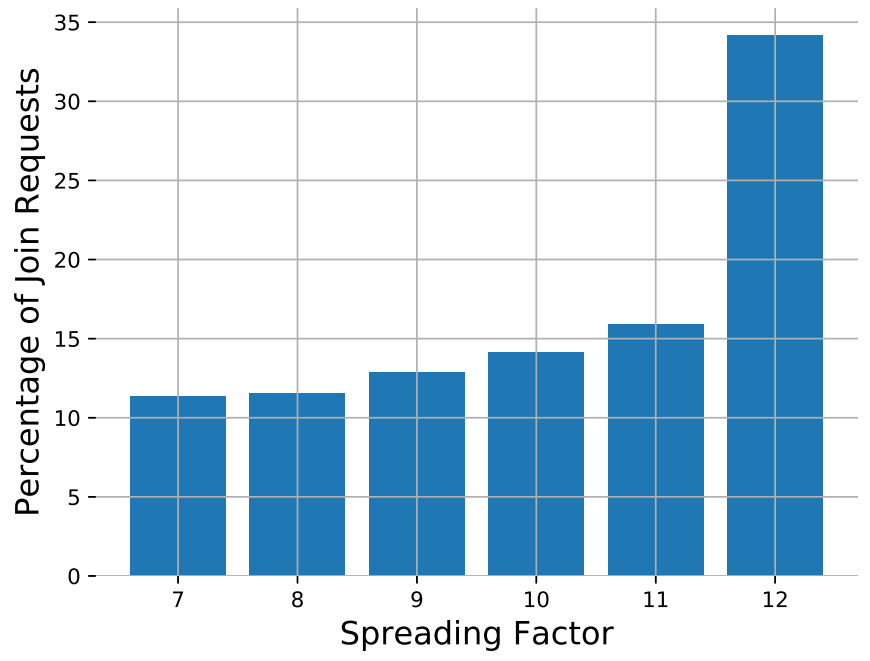

**Figure 12: Assessing the radio environment: Distribution of Spreading Factors for transmitted *JoinRequests* with a limited area populated by a high density of devices belonging to interfering actors $A$, $B$ and $D$.**

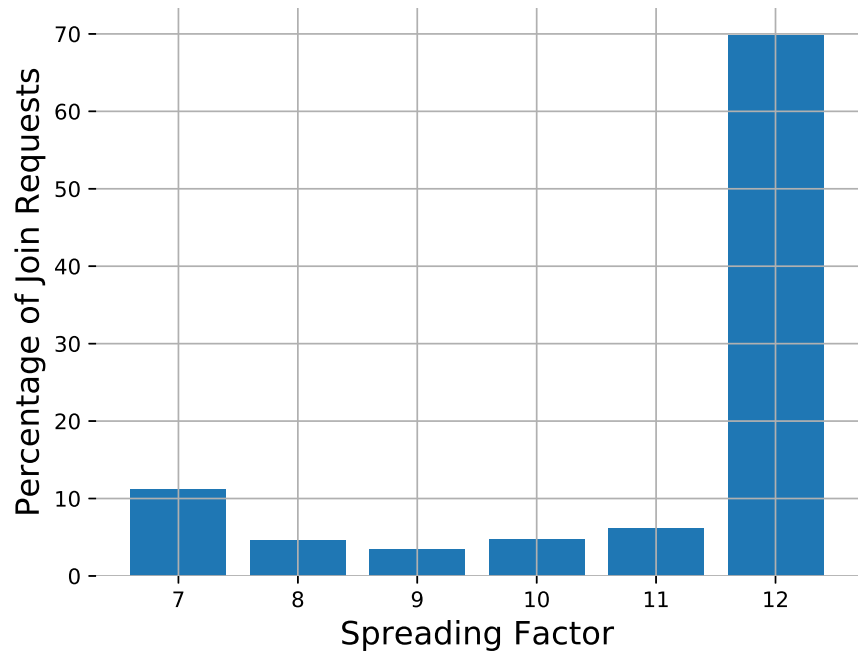

**Figure 14: Assessing the radio environment: Distribution of Spreading Factors for transmitted JoinRequests from 4 interfering actors $A$, $B$, $C$ and $D$.**

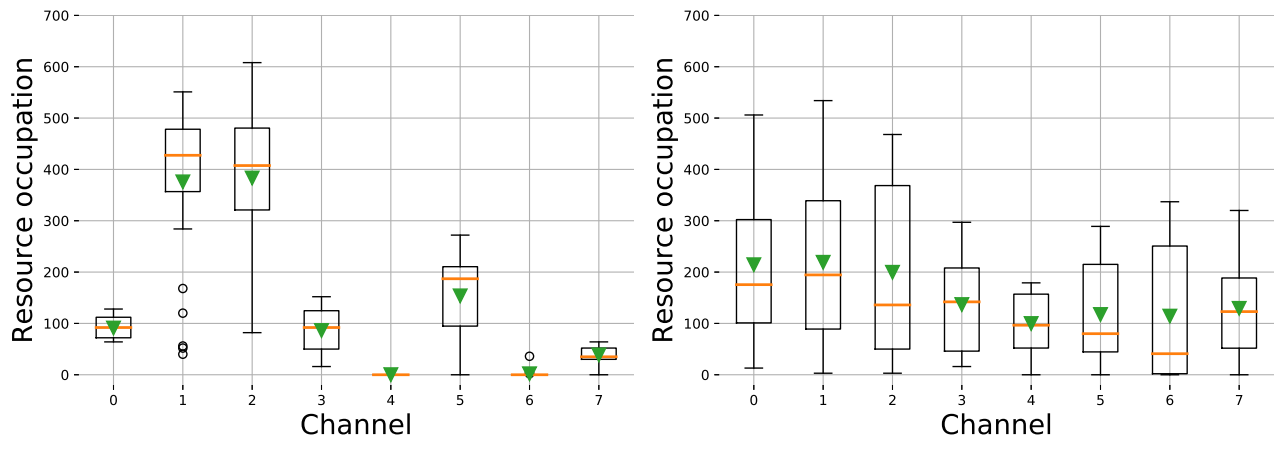

**Figure 13: Distribution of channel occupation: For gateway $D$ handling 63 end-devices non-cooperatively (a) and when handling 115 end-devices via *FLIP* (b).**

***Cooperation maximizes packet delivery ratio:*** Figure 11 shows the PDR for uncoordinated networks (top) and coordinated networks (bottom). In both cases, all devices are able to join the infrastructure. When $A$, $B$ and $C$ do not cooperate through *FLIP*, the end-devices are forced to adopt a higher SF to successfully join their owning gateway, which results in a lower average PDR, with a mean value of 90%. When $A$, $B$ and $D$ work together in *FLIP*, they determine the optimal gateway for each incoming end device, based not only on their own channel occupation, but also that of the other actors. The proportion of delegated vs directly managed end devices is shown by the dashed black line. Collaborative load balancing leads to a significant improvement in the overall PDR. Once the network reaches stability at the 16 hour mark, PDR climbs to a mean of 97% for $D$ and 100% for $A$ and $B$ when all devices have joined.

***Coordination requires time to stabilise:*** PDR naturally reaches stability earlier when actors do not collaborate, while under *FLIP* the stable state appears after the sixteen hour mark. This is due to the load balancing mechanism of FLIP waiting until the second join cycle of a device to reassign some for better conditions under a different actor. This behaviour can be observed for gateways $B$ and $D$ in Figure 11, where the share of delegated handling changes around the 8th hour (first join cycle) and the 16th hour (second join cycle) to improve upon bad radio conditions which were leading to a decrease in their respective PDR.

***FLIP balances channel occupation:*** For a more detailed view of how FLIP balances load, we assess channel occupation for gateway $D$ when operating uncooperatively, as shown in Figure 13a, and cooperatively, as shown in Figure 13b. Poor management of channel occupation can increase the percentage of failed join operations or saturate a given channel while leaving the others sub-optimally used, eventually affecting the PDR of the network. As can be seen from Figure 13a, when the 3 actors work as independent interfering networks, the overall channel occupation is ex-tremely uneven, with a concentration of handled end-devices on channels 3 and 5 and under-utilization of channels 4, 6 and 7 . In contrast, when *D* joins the *FLIP* federation, the load tend to be more evenly distributed among channels with a close median utilization as shown in Figure 13b. FLIP achieves this for an overloaded actor handling more than 70% of the deployed end-devices, where it reduces the average load of the most occupied channel by 30%.

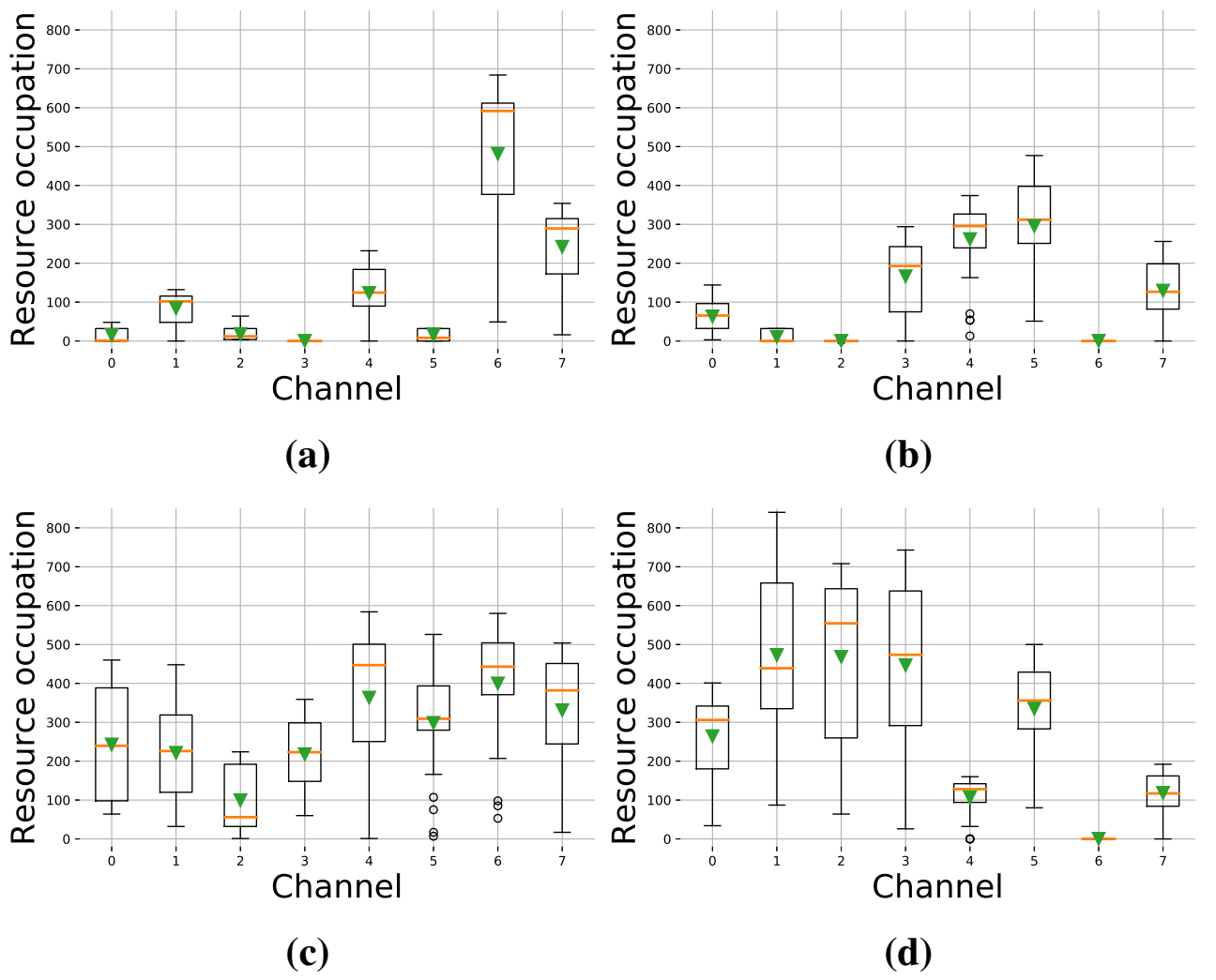

**Figure 15: Channel occupation: No collaboration between actors leads to unbalanced channel occupation for gateways *B* (a) and *C* (b). When working together, their combined occupation (c) tends to be flatten while absorbing the uncoordinated channel occupation spikes (d) produced by non-participating gateways *A* and *D*.**

Considered in sum, this experiment demonstrates that cooperation through the *FLIP* federation increases network reliability in comparison to uncoordinated gateways and makes more consistent and effective use of available gateway channels, through continuous collaborative optimization.

## 5.4 Handling Devices with Rising Contention

This last scenario tests the capability of *FLIP* to combat contention through increased collaboration between colocated gateways. This is accomplished by extending the previous setup of 3 actors with a fourth gateway, *C*, owning 100 supplementary end-devices. Naturally, this increases contention and makes it more difficult for end-devices to join the infrastructure.

The new end-devices are redeployed among the population in range of all four actors, in the same harsh radio conditions that were described previously. When the four actors do not collaborate, the resulting radio environment is suboptimal. The end-devices struggle to be handled due to their positions in houses, cellars and warehouses, and tend to succeed after many tries, having exhausted their lower *SpreadingFactors* according to LoRaWAN specifications [17]. This leads to a biased distribution of *SpreadingFactors*, as depicted in Figure 14, with 70% of devices joining at SF12. This vastly increases the chance of packet collisions and prevents nodes from joining the network.

***FLIP optimizes Spreading Factor allocation:*** The progressive collaboration of gateways significantly optimizes the allocation of spreading factors, with the number of nodes joining at SF12, falling from 70% as shown in Figure 14 to just 35% as shown in Figure 16. This leads to a commensurate reduction in time-on-air and thus contention.

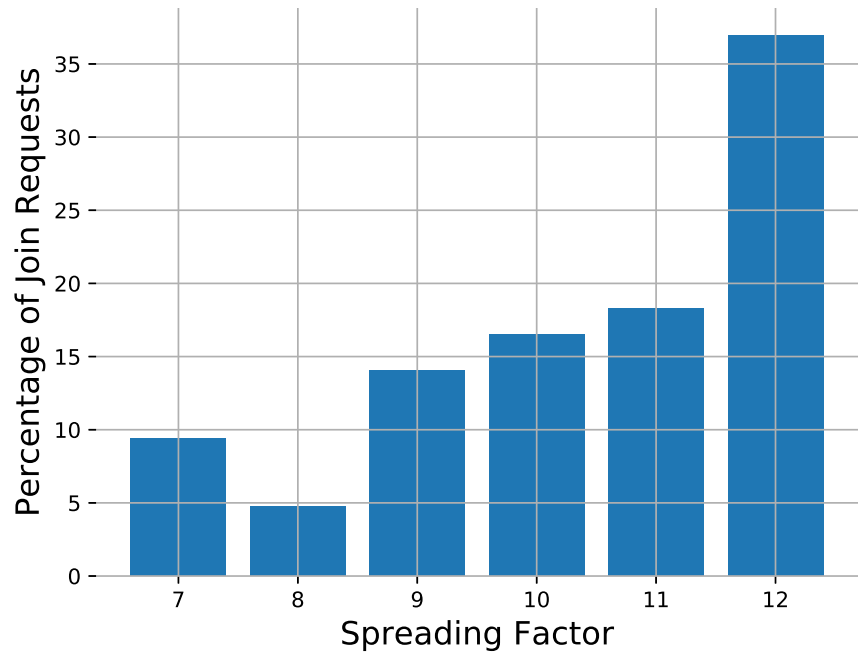

**Figure 16: Mitigating contention benefits: *FLIP* results in a reduced proportion of nodes joining at high SFs, diminishing radio resources consumption, increasing PDR and thus enabling more devices to join.**

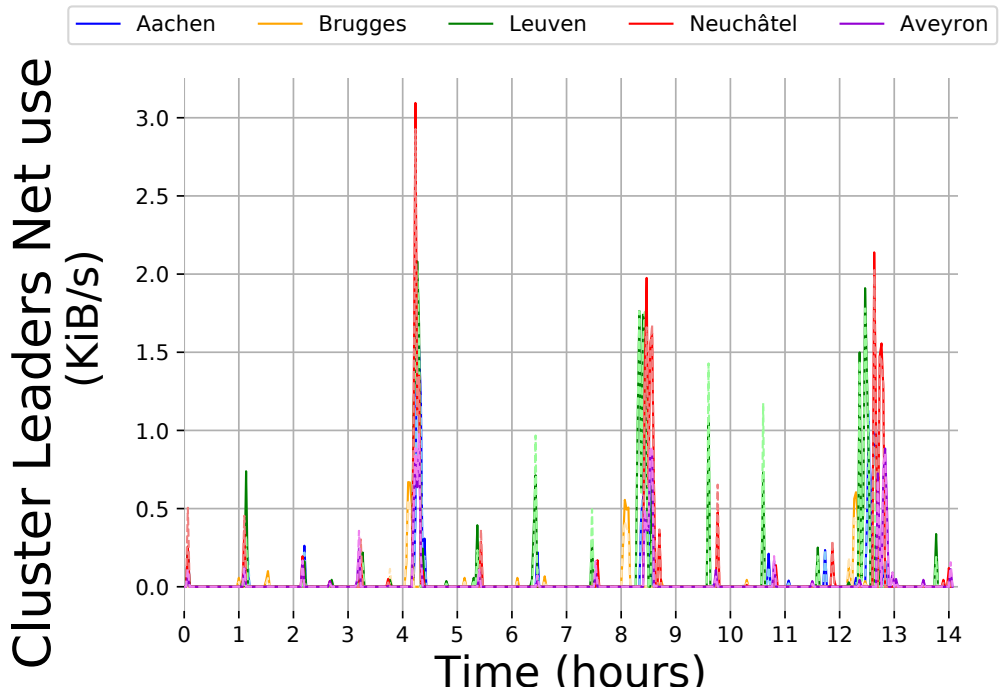

**Figure 17: Cost of *FLIP*: Inter-cluster network traffic, the cost of cooperation for advertising and handle remote deployed end-devices.**

***FLIP smooths channel utilization:*** As shown in Figure 15.a and Figure 15.b, the lack of collaboration between actors leads to unbalanced channel occupation for gateways *B* and *C* respectively. When these gateways move from competition to collaboration as shown in Figure 15.c and Figure 15.d, channel utilisation becomes more even, eliminating the channel occupation spikes and more effectively spreading network load thanks to collaboration handled by different actors.

Considered in sum, *FLIP* leads to a more efficient redistribution of network load, which lowers the overall channel occupation by more than 45%, thus allowing an increase of more than 20% in the number of end-devices that are able to join and participate in the infrastructure.

## 5.5 Scalability of *FLIP*

In this section, we evaluate the scalability of *FLIP* with respect to intra-cluster and inter-cluster communication.

Section 5.2 to 5.4 have showcased the benefits of *FLIP* in a range of network scenarios. These mechanisms are underpinned by a range of intra-cluster communication mechanisms that allow gateways to coordinates themselves and maintain their self-organized overlays. This section identifies the costs associated with these communication mecha-nisms.

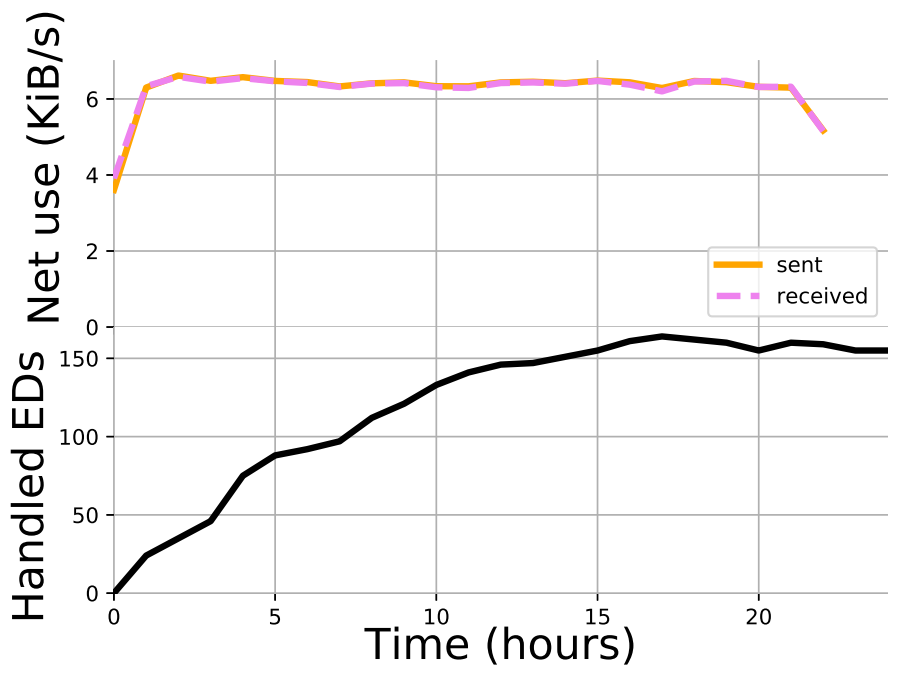

**Figure 18: Cost of *FLIP*: Inside a cluster, the network cost of cooperation does not vary with the increase in the load, staying constant for** 10 **as for** 150 **devices.**

To analyse the inter-cluster activity, we observe the network traffic between each leader, in a scenario where devices are deployed only in *Leuven* cluster and all belong to owners dispatched in the other 4 clusters. In this full roaming scenario we observe in Figure 17 each cluster disseminating its ownership markers to the others every 4 hours and the corresponding internal propagation of the received ones, the regular *kNN* cycles every hour for each involved leader. The traffic bumps observed during *Leuven*'s *kNN* mirror the leader participation in the self-stabilising overlays after first receiving ownerships markers, witnessed here growing and fading after having integrated the joining end-devices and reached converged state.

In fact, the generated inter-cluster traffic for the progressively deployed 150 devices is so small as to be negligible.

To analyse the intra-cluster scalability of *FLIP*, we observe the evolution of network activity over time for the scenario described in Section 5.4. With all gateways in range of each others, the similarity graph is not partitioned leading each gateway to have a similar network consumption profile as shown in Figure 18 for gateway *C*. During the first hour of the run, we note an initial increase in both transmitted and received data, corresponding to the initialisation phase and the first *kNN* round, which leads to the first establishment of the overlay that will cycle every hour, along with the cyclic ownership markers dissemination. As the number of handled end-devices increases as shown in Figure 18 along with the corresponding number of distributed consensus operations, we note no increase in the amount of data exchanged.

In fact, network resource consumption is indistinguishable for 10 handled end-devices vs 150, validating the scalability of the *FLIP* intra-cluster collaboration mechanisms.

# 6 RELATED WORK

This section reviews key work that is related to *FLIP*. Section 6.1 provides a brief overview of prior work studying the scalability of LoRaWAN. Section 6.2 then reviews related work on improving LoRaWAN scalability.

## 6.1 Studies of LoRaWAN Scalability

The scalability problems of LoRa are the subject of a growing body of work. Ramachandran [22] et al. built a plug and play sensor system using LoRaWAN and studied its performance in multiple suburban environments. Their evaluation revealed unpredictable performance that fell well below LoRaWAN marketing claims.

Bankov et al. [3] evaluate the channel access mechanisms of LoRaWAN in a typical EU LoRa configuration and show high packet error rates and packet loss rates, which dramatically limits either the number of devices per gateway or the rate at which messages can be transmitted. Adelantado et al. [2] examine the scalability of LoRa and show that that contention limits network capacity to an even greater degree than regulator duty cycles. The authors find that in fully utilized LoRaWAN networks, the probably of successful packet transmission drops below 15%.

Reynders et al. [24] investigated the efficacy of the Adaptive Data Rate (ADR) protocol that is used in LoRaWAN and found that it is insufficient to cope with interference from co-located third-party networks. Moreover, MAC commands require upstream acknowledgements, which themselves compete for the shared channel, reducing the uplink throughput, which research has shown can can lead to congestion collapse [21].

The studies above are unanimous in their assessment that the current LoRaWAN protocol is inadequate to support large-scale and dense deployments of end-devices, and yet the global roll-out of LoRaWAN continues. This shows an urgent need for systems such as *FLIP*that can deliver practical scalability improvements for today's LoRa devices.

## 6.2 Improvements to LoRaWAN

Ifikhar et al. [12] also identify the difficulty of tailoring LoRaWAN configurations in the face of unpredictable surrounding environments. The authors introduce a new MAC protocol, DeltaIoT, to tackle this problem by applying the MAPE-K (Monitior, Analyse, Plan, Execute, Knowledge) methodology to guide protocol behaviour. The key difference between DeltaIoT and *FLIP* is that DeltaIoT is not backwards compatible with the millions of LoRa end devices that are already deployed in the field.

Voigt et al. [26] study the problem of inter-network interference for LoRa networks through extensive simulation and show that interference arising due to co-located networks acting in an uncoordinated fashion can radically reduce the performance of a LoRa network. The authors also demonstrate that this effect can be ameliorated by deploying additional base-stations equipped with directional antennae in order to reduce the scope for contention. This approach is clearly more complex and costly for end-users, which is why *FLIP* addresses the root of the problem; that co-located networks compete rather than cooperate.

*FLIP* is not the first proposal for a distributed LoRaWAN management solution. Worldline [4] introduced BcWAN, an architecture for LoRaWAN based on a blockchain. In contrast to our work BcWAN does not consider contention, focussing instead on providing secure and decentralized roaming across a LoRaWAN federation using a blockchain-based security mechanism. This form of decentralised security is complementary to, and compatible with *FLIP*.

## 7 Future work

The first priority of our future work is to undertake a complete security analysis of *FLIP* in order to validate the security guarantees that it offers and assess the potential impact of malicious gateways that engage in behavior such as: leader hijacking, the injection of falsified data or ownership impersonation. Developing countermeasures that address these problems is a core requirement to promote the mass adoption of *FLIP*.

We are also actively working on the problem of *free riding*, wherein some federation members may consume far more resources than they contribute. This is a well known problem in peer-to-peer systems and a variety of techniques exist to encourage participant reciprocity. We are currently investigating which of these schemes are most suitable for the *FLIP* ecosystem.

## 8 CONCLUSION

This paper introduced *FLIP*, a novel approach to reducing contention on LoRaWAN interfering networks through the formation of a federation of gateways. Gateways in the *FLIP* federation cooperate to reduce localised contention and balance load across the LoRa network while offering by default transparent roaming abilities. The *FLIP* architecture is fully distributed and emerges autonomously based upon consensus-driven and localised decision making. Any gateway owner may participate in the *FLIP* federation, by simply installing the *FLIP* client software on their gateway.

We evaluated *FLIP* in a real world deployment that is comprised of *417* LoRaWAN end devices and 17 gateways deployed across Europe. Our evaluation reveals that *FLIP* gateways significantly outperform independent gateways: we observed a x2 gain packet delivery ratio up to 100%, 20% increase in proportion of end devices that join successfully and better load balancing on radio resource consumption. We studied the performance of both on local and global levels, and demonstrated the benefits our federation embodies while preserving full deployment freedom and a completely distributed architecture. *FLIP* achieves these advances while requiring no changes to the firmware of current LoRaWAN end devices and remaining hardware compatible with all standard LoRa gateways. By offering a free and open federation for IoT end-users of any size, *FLIP* empowers its users, while shielding them from the costs and complexities of building a scalable LoRaWAN backhaul infrastructure.

In keeping with the open philosophy of *FLIP*, all software, hardware designs and experimental data described in this paper are available in the official repository [7].

## 9 References

[1] N. Abramson. The throughput of packet broadcasting channels. *IEEE Transactions on Communications*, 25(1):117–128, January 1977.

[2] F. Adelantado, X. Vilajosana, P. Tuset-Peiró, B. Martínez, and J. Melià. Understanding the limits of LoRaWAN. *CoRR*, abs/1607.08011, 2016.

[3] D. Bankov, E. Khorov, and A. Lyakhov. On the limits of LoRaWAN channel access. In *2016 International Conference on Engineering and Telecommunication (EnT)*, pages 10–14, Nov 2016.

[4] M. Bezahaf, G. Cathelain, and T. Ducrocq. Bcwan: A federated low-power wan for the internet of things (industry track). In *Proceedings of the 19th International Middleware Conference Industry*, Middleware '18, pages 54–60, New York, NY, USA, 2018. ACM.

[5] M. C. Bor, U. Roedig, T. Voigt, and J. M. Alonso. Do LoRa low-power wide-area networks scale? In *Proceedings of the 19th ACM International Conference on Modeling, Analysis and Simulation of Wireless and Mobile Systems*, pages 59–67. ACM, 2016.

[6] Dave Kjendal. LPWAN standards, advantages, and use cases, 2017.

[7] Delbruel et al. Federated Long-range low-power Iot Protocol (FLIP). Official repository at `https://gitlab.com/FLIP_federation/`, April 2019.

[8] M. D. Ekstrand, J. T. Riedl, J. A. Konstan, et al. Collaborative filtering recommender systems. *Foundations and Trends® in Human–Computer Interaction*, 4(2):81–173, 2011.

[9] O. Georgiou and U. Raza. Low Power Wide Area Network Analysis: Can LoRa Scale? *IEEE Wireless Communications Letters*, 2017.

[10] J. Haxhibeqiri, F. Van den Abeele, I. Moerman, and J. Hoebeke. LoRa scalability: A simulation model based on interference measurements. *Sensors*, 17(6):1193, 2017.

[11] S. M. Hedetniemi, S. T. Hedetniemi, and A. L. Liestman. A survey of gossiping and broadcasting in communication networks. *Networks*, 18(4):319–349, 1988.

[12] M. U. Iftikhar, G. S. Ramachandran, P. Bollansée, D. Weyns, and D. Hughes. Deltaiot: A self-adaptive internet of things exemplar. In *Proceedings of the 12th International Symposium on Software Engineering for Adaptive and Self-Managing Systems*, SEAMS '17, pages 76–82, Piscataway, NJ, USA, 2017. IEEE Press.

[13] M. Jelasity, R. Guerraoui, A.-M. Kermarrec, and M. van Steen. The peer sampling service: Experimental evaluation of unstructured gossip-based implementations. In *Proceedings of the 5th ACM/IFIP/USENIX International Conference on Middleware*, Middleware '04, pages 79–98, New York, NY, USA, 2004. Springer-Verlag New York, Inc.

[14] M. Jelasity, S. Voulgaris, R. Guerraoui, A.-M. Kermarrec, and M. van Steen. Gossip-based peer sampling. *ACM Trans. Comput. Syst.*, 25(3), Aug. 2007.

[15] B. M. Kapron, D. Kempe, V. King, J. Saia, and V. Sanwalani. Fast asynchronous byzantine agreement and leader election with full information. *ACM Trans. Algorithms*, 6(4):68:1–68:28, Sept. 2010.

[16] J. C. Liando, A. Gamage, A. W. Tengourtius, and M. Li. Known and unknown facts of LoRa: Experiences from a large-scale measurement study. *ACM Transactions on Sensor Networks (TOSN)*, 15(2):16, 2019.

[17] LoRa Alliance. LoRaWAN Specifications 1.0, 2015.

[18] LoRa Alliance. LoRaWAN Specifications 1.1, Oct. 2017.

[19] J. M. Marais, R. Malekian, and A. M. Abu-Mahfouz. Evaluating the LoRaWANprotocol using a permanent outdoor testbed. *IEEE Sensors Journal*, 19(12):4726–4733, June 2019.

[20] K. Mikhaylov, J. Petäjäjärvi, and J. Janhunen. On LoRaWAN scalability: Empirical evaluation of susceptibility to inter-network interference. In *2017 European Conference on Networks and Communications (EuCNC)*, pages 1–6. IEEE, 2017.

[21] A. Pop, U. Raza, P. Kulkarni, and M. Sooriyabandara. Does bidirectional traffic do more harm than good in LoRaWAN based LPWA networks? *CoRR*, abs/1704.04174, 2017.

[22] G. Ramachandran, F. Yang, P. Lawrence, S. Michiels, W. Joosen, and D. Hughes. µpnP-WAN: Experiences with LoRa and its deployment in DR Congo. pages 63–70, 6 2017.

[23] U. Raza, P. Kulkarni, and M. Sooriyabandara. Low power wide area networks: An overview. *IEEE Communications Surveys & Tutorials*, 19(2):855–873, 2017.

[24] B. Reynders, W. Meert, and S. Pollin. Range and coexistence analysis of long range unlicensed communication. In *2016 23rd International Conference on Telecommunications (ICT)*, pages 1–6, 2016.

[25] B. Robbe and W. Danny. A QoS-Aware adaptive mobility handling approach for LoRa-based IoT systems. In *2018 IEEE 12th International Conference on Self-Adaptive and Self-Organizing Systems (SASO)*, pages 130–139, Sep. 2018.

[26] T. Voigt, M. Bor, U. Roedig, and J. Alonso. Mitigating inter-network interference in LoRa networks. In *Proceedings of the 2017 International Conference on Embedded Wireless Systems and Networks*, EWSN '17, pages 323–328, USA, 2017. Junction Publishing.